\documentstyle[epsfig,emulateapj]{article}

\begin{document}

\newcommand{\Mo}{\mbox{$\rm M_\odot$}}
\newcommand{\Lo}{\mbox{$\rm L_\odot$}}
\newcommand{\mic}{\mbox{$\rm \mu m$}}
\newcommand{\ivol}{\mbox{$\rm cm^{-3}$}}
\newcommand{\isup}{\mbox{$\rm cm^{-2}$}}
\newcommand{\isec}{\mbox{s$^{-1}$}}
\newcommand{\Av}{\mbox{$A_{\rm V}$}}
\newcommand{\Ne}{\mbox{$N_{\rm e}$}}
\newcommand{\Np}{\mbox{$N_{\rm p}$}}
\newcommand{\Te}{\mbox{$T_{\rm e}$}}
\newcommand{\ten}[1]{\mbox{$10^{#1}$}}
\newcommand{\xten}[1]{\mbox{$\times 10^{#1}$}}
\newcommand{\wl}{\mbox{$\lambda$}}
\newcommand{\forb}[2]{\mbox{$[{\rm #1\, #2}]$}}
\newcommand{\Ha}{\mbox{H$\alpha$}}
\newcommand{\PA}{\mbox{Pa$\alpha$}}
\newcommand{\Hb}{\mbox{H$\beta$}}
\newcommand{\QH}{\mbox{$Q(\rm H)$}}

\lefthead{Schreier et al.}
\righthead{Disk in Centaurus A}

\title{Evidence for a 20pc disk at the nucleus of Centaurus A
\footnote{Based on
observations with the NASA/ESA Hubble Space Telescope, obtained at the
Space Telescope Science Institute, which is operated by AURA, Inc.,
under NASA contract NAS 5-26555 and by STScI grant GO-3594.01-91A}}

\author{Ethan J. Schreier, Alessandro Marconi\altaffilmark{1,2},
David J. Axon\altaffilmark{3,4}, Nicola Caon,
Duccio Macchetto\altaffilmark{3}}

\affil{Space Telescope Science Institute \\
       3700 San Martin Drive, Baltimore, MD 21218, USA\\}

\author{Alessandro Capetti}
\affil{Osservatorio Astronomico di Torino\\
		 Strada Osservatorio 20, I-10025 Pino Torinese, ITALY\\} 

\author{James H. Hough, Stuart Young}
\affil{Division of Physics and Astronomy, 
Department of Physical Sciences, University of Hertfordshire\\
College Lane, Hatfield, Herts AL10 9AB, UK\\}

\author{Chris Packham}
\affil{Isaac Newton Group, Sea Level Office\\Apartado de Correos, 321, 
38780 Santa Cruz de La Palma,\\Islas Canarias, SPAIN\\}

\altaffiltext{1}{Osservatorio Astrofisico di Arcetri,
		 Largo E. Fermi 5,
		 50125 Firenze,
		 ITALY}
\altaffiltext{2}{Dipartimento di Astronomia e Scienza dello Spazio,
		 Universit\`a di Firenze,
		 Largo E. Fermi 5,
		 50125 Firenze,
		 ITALY}
\altaffiltext{3}{Associated with Astrophysics Division, Space Science Dept.,
ESA}
 
\altaffiltext{4}{On leave from Nuffield Radio Astronomy Laboratory,
Jodrell Bank, Univ. of Manchester, UK}

\authoremail{schreier@stsci.edu}

\begin{abstract}
We report HST NICMOS observations of the central region of NGC 5128
at 2.2\mic\ and in \PA.
The continuum images show extended emission typical of 
an elliptical galaxy and a strong unresolved central source we identify  as 
the nucleus of the galaxy. Its position is consistent with ground-based IR 
and radio data, and with the peak of reddening found with 
WF/PC-1. In \PA, we detect a prominent elongated structure, centered on the 
nucleus, extended by $\simeq2$\arcsec at a position angle of $\simeq$33\arcdeg,
and with a major to minor axis ratio of $\sim2$.
We interpret this as an inclined, $\sim 40$ parsec diameter, thin nuclear
disk of ionized gas rather than  a jet-gas cloud interaction. We
do see several weaker \PA\ features, some of which may be circumnuclear 
gas clouds shocked by the X-ray/radio jet.  
The disk is one of the smallest ever observed at the nucleus of an AGN.  
It is not perpendicular to the jet, but consistent with being oriented 
along the major axis of the bulge.  If it represents the warped outer 
portion of an accretion disk around a black hole, we conclude that 
even on the scale of a few parsecs, the disk is dominated by the galaxy 
gravitational potential and not directly related to the
symmetry axis of the AGN.

\end{abstract} 


\section{Introduction}

NGC 5128 (Centaurus A), the nearby giant elliptical galaxy, 
hosts the closest powerful active galactic nucleus.  This double-lobed radio 
source contains a strong jet discovered in X-rays 
(\cite{schreier:79}, \cite{feigelson:81}) and well-studied in radio
with the VLA 
(\cite{schreier:81}, \cite{burns:83}, \cite{clarke:86}) and with the VLBI 
(cf. \cite{jones:96}).
The proximity of NGC 5128 makes it one of the best candidates for 
studying the inner region around a massive black hole, assumed by the 
standard model to be at the core of all AGNs
(e.g. \cite{blandford:91}, \cite{antonucci:93}).
However, the large 
warped dust lane, approximately an arcminute wide with associated gas, 
young stars, and HII regions, dominates the morphology and polarization 
properties at visible wavelengths, effectively obscuring the nucleus and 
inner half-kiloparsec of the galaxy and the jet.  

Ground-based IR observations have provided evidence for a strong, heavily 
polarized source at the nucleus (cf. \cite{bailey:86}, \cite{packham:96}).
Previous HST WF/PC-1 imaging polarimetry of the inner region of NGC5128 
(\cite{schreier:96}) identified the nucleus of the galaxy with the most 
obscured and reddened feature of the emission, several arcseconds SW of 
the brightest optical emission near the center of the dust lane and 
consistent with the IR results.  
 
While the discovery of gas disks at the centers of nearby galaxies with HST 
has opened up new possibilities for determining masses of central black holes 
by studying kinematics of the disks (cf. \cite{harms:94}, \cite{ferrarese:96},
\cite{macchetto:97},  \cite{bower:98}) the dust lane does 
not permit such studies for NGC 5128 
in the optical.  We report here HST Near Infrared Camera and Multi-Object 
Spectrometer (NICMOS) observations at 2.2\mic\ and \PA\ which allow 
us, for the first time, to study the structure of the circumnuclear region at 
high resolution.  We summarize the observations and data reduction in Section 
2, the results in Section 3, and discuss the implications of these data for standard AGN and jet models in Sections 4 and 5.  Throughout, we assume a distance to Centaurus A of 3.5 Mpc
(\cite{hui:93}), whence 1\arcsec$\simeq$17pc.

\section{Observations and Data Reduction}

The nuclear region of NGC 5128 was observed for three orbits on 11 August 
1997 using NICMOS Camera 2 (0\farcs075/pixel) with F222M, F187N and 
F190N filters. All observations 
were carried out with a MULTIACCUM sequence (\cite{mackenty:97}), 
the detector read out non-destructively several times during each 
integration to facilitate removal of cosmic rays and saturated pixels.
The integration times were 2304 sec for the F187N and F190N filters
(\PA\ and adjacent continuum) and 1280 sec for the F222M filter (2.2\mic\
medium band continuum). The last observation was also performed
off-source to enable telescope thermal background subtraction. The data were calibrated using the pipeline software CALNICA v3.0 
(\cite{bushouse:97}) and the best reference files in the Hubble Data 
Archive to produce flux calibrated images.

A continuum-subtracted \PA\ image was obtained by direct subtraction of the 
F187N and F190N images.  We verified the continuum subtraction by rescaling 
the continuum by up to $\pm10\%$ before subtraction and establishing that 
this did not significantly affect the observed emission line structure.  

A small (few percent) drift in the NICMOS bias level resulted in spatially dependent residuals in the calibrated images (the ``pedestal'' problem, 
\cite{pedestal}).  This effect was removed by fitting and subtracting a 
first degree polynomial surface to each quadrant of the continuum subtracted 
images.

\section{\label{sec:results} Results}

\begin{figure}
\centering
\epsfig{figure=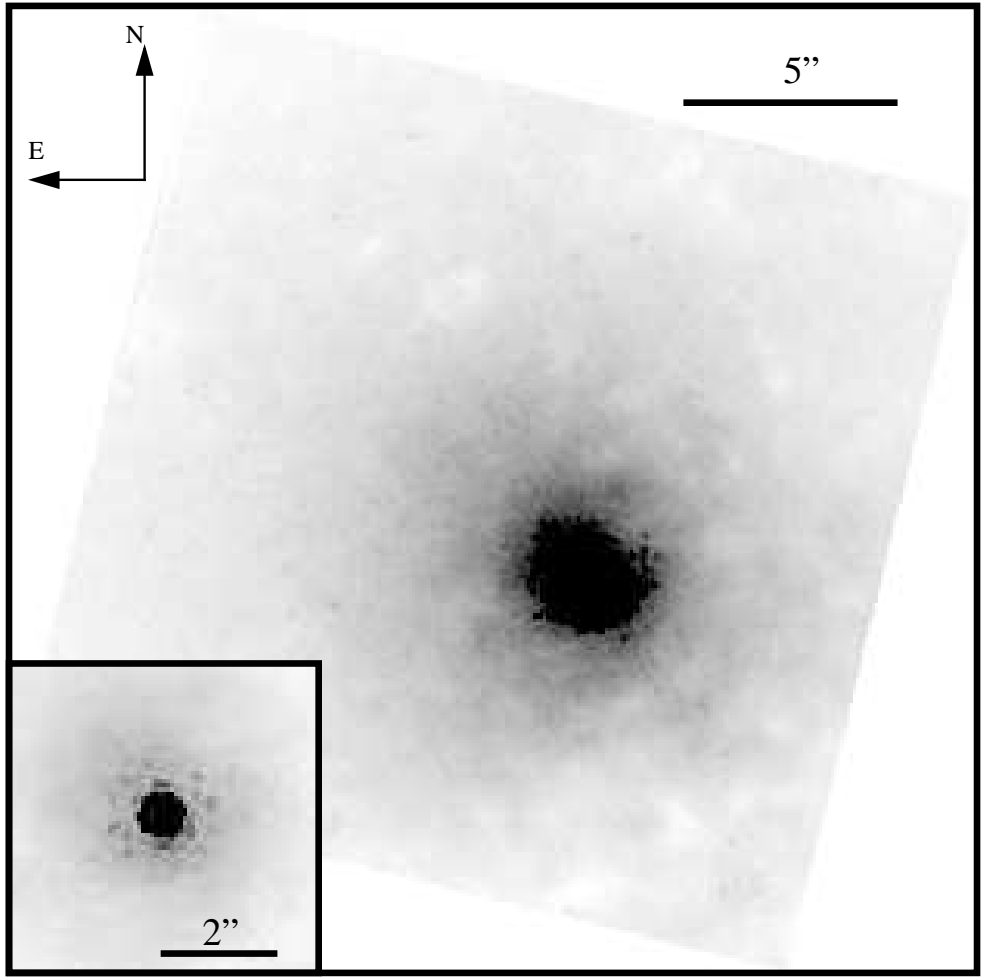,width=0.7\linewidth}
\vskip 0.5cm
\caption{\label{fig:continua} Gray scale F222M image
of the nuclear region; dynamic range is 0.45--6.2
(units of \ten{-16} erg cm$^{-2}$ s$^{-1}$ \AA$^{-1}$ arcsec$^{-2}$).
Inset shows the region
around the unresolved source identified as the nucleus;
dynamic range is 3.6--14 (same units).
Dots and other linear structures are artifacts
of the NICMOS PSF.} 
\end{figure}

The calibrated 2.2 $\mu$m continuum image in Fig. \ref{fig:continua} shows  
the rather smooth and regular structure expected for a typical 
elliptical galaxy.  This smooth structure on small scales contrasts strongly 
with the patchy and irregular emission observed on the same spatial scales in 
R and I with WF/PC-1 (\cite{schreier:96}).  The continuum image is dominated by 
a strong unresolved source which we interpret as the nucleus of the galaxy.  

\begin{figure}
\centering
\epsfig{file=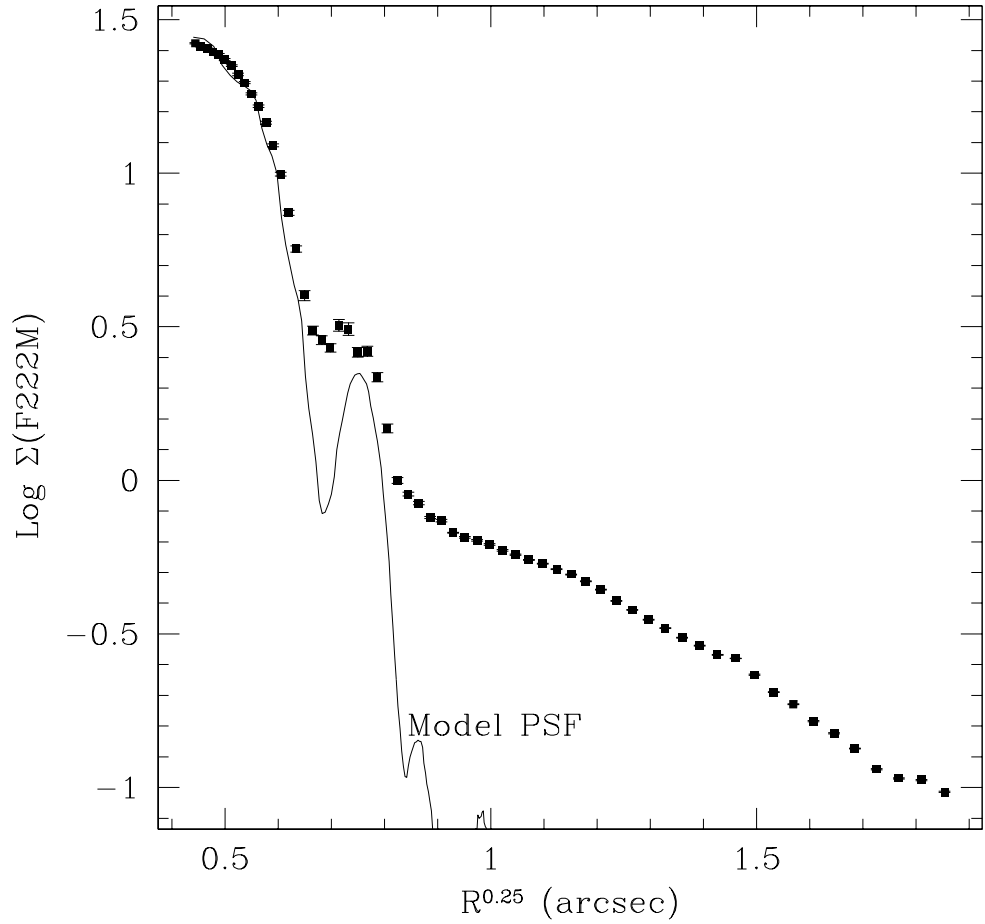,width=0.7\linewidth}
\vskip 0.5cm
\caption{\label{fig:radprof} Radial light distribution in F222M 
(filled squares). The logarithm of the azimuthally averaged
surface brightness derived from ellipse fitting (units
of \ten{-15} erg cm$^{-2}$ s$^{-1}$ \AA$^{-1}$ arcsec$^{-2}$)
is plotted as a function of the quartic root of the distance
from the nucleus. Error bars are comparable to or smaller than the squares.
Solid line is the best-fit model PSF.}
\end{figure}

The intensity of the nucleus is estimated by fitting a model PSF
derived using the TinyTim software (\cite{tinytim}). 
 We took particular care in removing
the PSF artifacts present beyond the first Airy ring.
The total non-stellar plus stellar flux of the unresolved nuclear source 
is (2.3$\pm$0.1)\xten{-15} erg cm$^{-2}$ s$^{-1}$ \AA$^{-1}$ (magnitude 
10.52$\pm$0.05 in the Vega reference system).
Fig. \ref{fig:radprof} 
shows the azimuthally averaged radial light profile obtained by 
fitting ellipses to the isophotes in the F222M image and comparing 
with the model PSF fitted to the data.

The nucleus position is RA=13:25:27.46, Dec=-43:01:10.2 (J2000), with the 
$\pm$1\arcsec\ uncertainty of the HST Guide Star Catalogue (GSC).  This 
position is consistent with that found in the radio and
ground based IR data, and with both the peak of reddening
and the peak of polarization found with WF/PC-1.

\begin{figure}
\centering
\epsfig{file=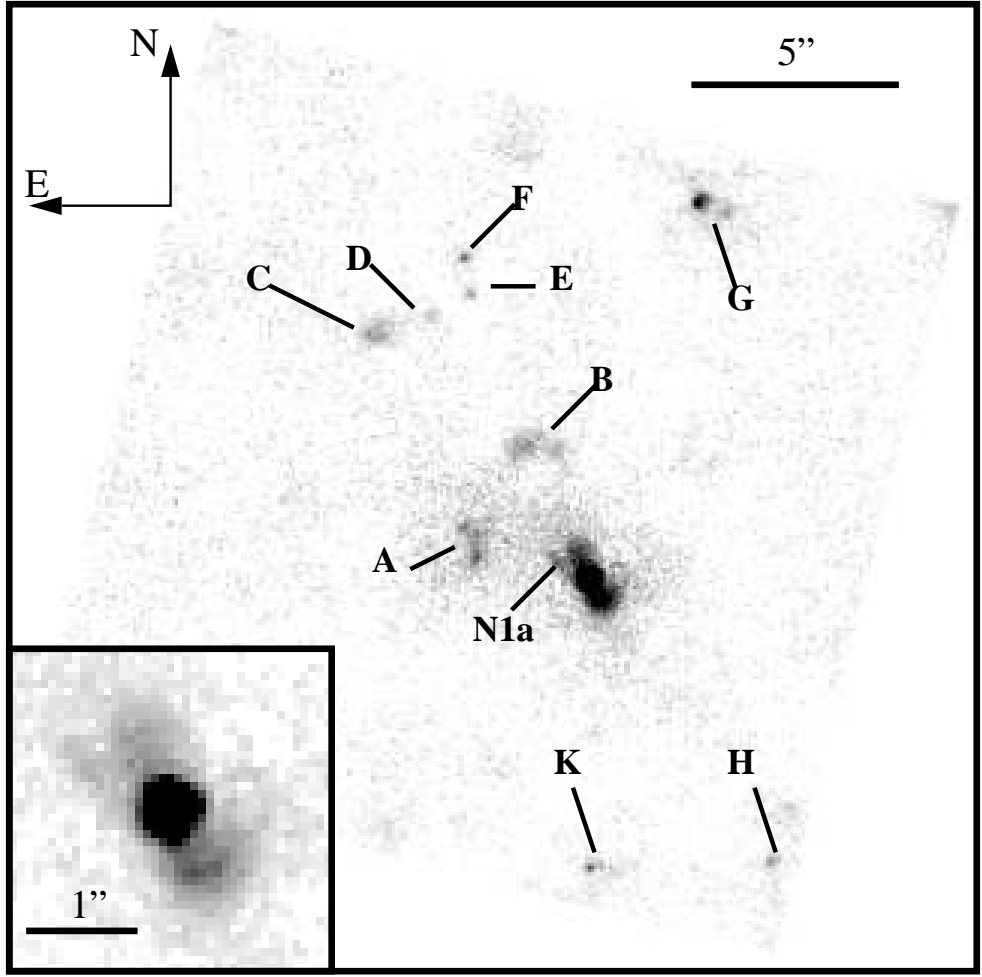,width=0.7\linewidth}
\vskip 0.5cm
\caption{\label{fig:pa} Grey scale continuum-subtracted \PA\ image;
dynamic range is 0--3.6 (units of \ten{-14} erg cm$^{-2}$ s$^{-1}$
arcsec$^{-2}$).
Inset shows the central region; dynamic range is 0.18--11 (same units).
The emission line knots and filaments are labelled for reference.}
\end{figure}

In contrast to the relatively smooth light distribution seen in F222M, the 
continuum subtracted \PA\ image (Fig. \ref{fig:pa}) reveals a number of 
emission features above the threshold of
6\xten{-15} erg cm$^{-2}$ s$^{-1}$ arcsec$^{-2}$ 
(3$\sigma$ of the noise measured in emission free 
areas).  In addition to the unresolved nuclear source, there is a prominent 
elongated structure, with some  
sub-structure, centered on the nucleus, 
a finger (N1a) $\simeq$0\farcs9 N-E of the nucleus,
two arc-like filaments (A,B) located,
respectively, at $\simeq$ 2\farcs9 N-E and 3\farcs5 N-N-E 
of the nucleus, and several other weaker emission knots and filaments 
distributed over the field of view.

The dominant feature by far is the structure around the nucleus.  Isophotal 
ellipse
fitting gives an average ellipticity of 0.5 and a constant position angle of 
33\arcdeg\ outside a radius of $\sim$0\farcs1,
within which the nucleus dominates.  The total extension 
along the major axis is $\simeq$ 2\farcs3 ($\simeq$ 40pc) and the total 
flux from the extended nuclear emission above the level
of 1.1\xten{-14} erg cm$^{-2}$ s$^{-1}$ arcsec$^{-2}$
is (3.9$\pm$0.4)\xten{-14} erg cm$^{-2}$ s$^{-1}$.
In comparison, the observed flux from the unresolved \PA\ emission is 
(5.7$\pm$0.6)\xten{-14} erg cm$^{-2}$ s$^{-1}$. 
The inset in Fig. \ref{fig:pacont} overlaying the \PA\ contours on 
a gray scale plot of the nuclear emission after subtracting the 
isophotal fit model
 clearly shows N1a and another sub-structure S-S-E of the nucleus. 

\section{Extended \PA\ Disk} 

The elongated structure around the nucleus seen in \PA\ emission has 
an overall extent of $\sim 40\times20$ parsecs.  Its position angle of 
33\arcdeg is approximately perpendicular to that of the dust lane.  
It is not oriented along the axis of the X-ray/radio jet, which has 
a position angle of $\sim$ 50\arcdeg-55\arcdeg\ as 
seen at X-ray (\cite{schreier:79}), cm (\cite{burns:83}) and VLBI 
(\cite{jones:96}) wavelengths.
It is also not perpendicular to the jet axis, 
as reported for an extended structure seen with ISOCAM at much lower 
resolution in the near IR at P.A. 
$\sim145\arcdeg$ (\cite{vigroux:97}), nor is it consistent with the 
orientation of a ring with a radius of $\simeq$100 pc perpendicular to the 
radio jet, reported by \cite{rydbeck:93}.  As such, we believe that we are 
observing a new feature, not directly related to the radio/X-ray jet or to 
any previously reported optical or IR features.

We have verified that the observed elongation is not the result of varying 
extinction in front of a smooth circular emission region. Indeed, if the 
extinction is maximum at the nucleus and decreases outwards in a direction 
perpendicular to the dust lane, the resulting morphology would be elongated 
in the direction of this extinction gradient.  We calculated the IR reddening 
correction map by adopting the procedure presented in \cite{schreier:96},  
comparing the colors obtained in recent WFPC2 observations 
(\cite{schreier:98}) with our NICMOS continuum images, using the 
reddening law of \cite{reddenlaw} with $R_V=3.1$,
and assuming a constant intrinsic 
color over the field of view of NICMOS; this latter assumption is supported 
by an I-K vs V-I color diagram where most
of the points lie tightly along the 
reddening line.  Applying the derived reddening correction 
to the \PA\ image does not significantly affect the
morphology of the observed structure.  

HST observations of other AGNs suggest three possible explanations for the 
observed elongated structure:
1) gas which is shocked and compressed by interaction with the jet;
2) illumination of
 gas clouds by an anisotropic nuclear radiation field 
(i.e. an ionization cone, cf. \cite{robinson:97}
and references therein);
and 3) a gaseous disk, such as that of M87 (\cite{harms:94}).

\begin{figure}
\centering
\epsfig{file=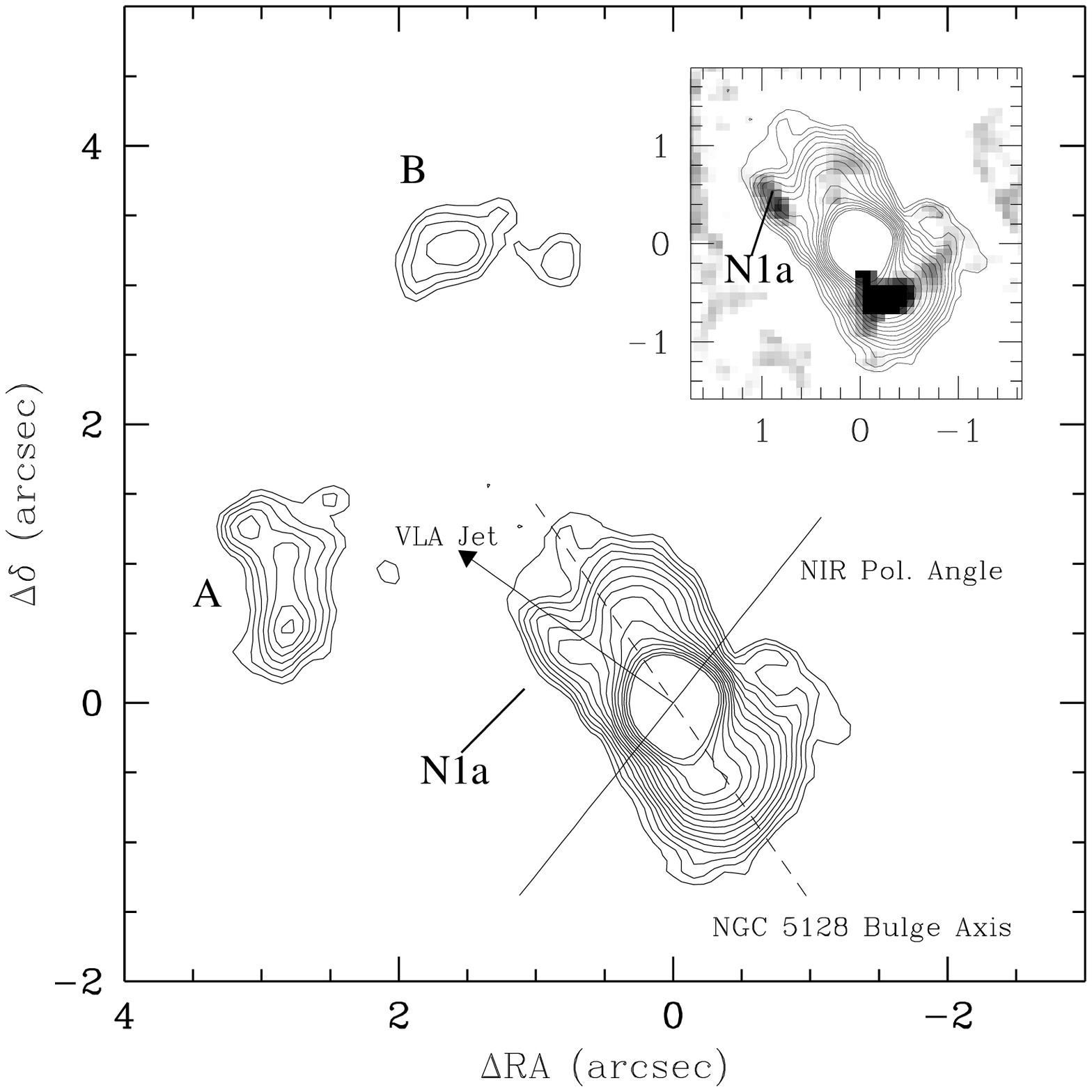,width=0.7\linewidth}
\vskip 0.5cm
\caption{\label{fig:pacont} Isophotes of the \PA\ emission near the 
nucleus. Contours have constant logarithmic spacing from 6.2 to 53 (units 
of \ten{-15} erg cm$^{-2}$ s$^{-1}$ arcsec$^{-2}$).  The origin is the nucleus position.
The arrow shows the PA of the VLA radio jet (\cite{clarke:86}), the solid 
line the K band polarization vector (2\farcs25 aperture on the IR peak, 
\cite{packham:96}), and the dashed line the position angle from V band 
isophotal ellipse fitting of the large scale NGC 5128 bulge (\cite{dufour:79}).
Gray scales in the inset are residuals obtained by subtracting the
fitted elliptical model; dynamic range is 2.2--11 (same units).}
\end{figure}

The structure's position angle differs by 
$\sim20\arcdeg$ from the jet axis
(see Fig. \ref{fig:pacont}).  This significant misalignment suggests the 
emission is not due to a current jet-cloud interaction, especially when we 
see other \PA\ features well correlated with the radio/X-ray jet morphology 
(see next section).   It has 
long been hypothesized that the radio jet had a different (smaller) position 
angle in the past, creating the N-S orientation of the outer lobes, and then 
rotating to form first the inner lobes and then the currently observed
X-ray/radio 
jet.  In this regard we note that the outer edge of the N-E inner lobe does 
have structure aligned with the position angle of our feature; both could be 
gas shocked by the jet (perhaps two-sided) at some recent time, and not yet 
cooled down.  Existing radio and X-ray measurements do not provide added 
information on this small spatial scale.

The $\sim20\arcdeg$ misalignment with the jet does allow the elongated 
structure to lie well within a putative ionization cone.  We note that 
most of the other detected emission features would also lie within
such a cone if it has an opening half-angle of at least $\sim30\arcdeg$,
well within the limits postulated by the unified model
(e.g. \cite{antonucci:93}). As it stands, 
we cannot rule out clouds densely distributed along this position angle,
embedded in a large opening angle radiation 
field.  However, the relative thickness of the feature in the transverse 
direction and the symmetry around the nucleus suggest
that this is not the case.

We find the simplest interpretation for the elongated emission to be 
a gaseous disk around the nucleus, as seen on larger scales around other 
galactic nuclei.  
It could be the outer part of an accretion disk, expected around 
a massive black-hole at the core of this AGN.  If the structure is indeed a 
thin circular disk, the axial 
ratio suggests an inclination of $\simeq$60\arcdeg\ .  Its radius of 
$\sim 1\farcs2$ ($\sim 20$pc) makes it significantly smaller 
than the hundred parsec scale stellar disks observed by HST 
at the centers of many galaxies.

This gas disk could be readily ionized by the powerful AGN, seen in X- and 
$\gamma$-rays.  The flux of photons required to keep the \PA\ emitting 
material ionized can be computed following \cite{osterbrock:89}:
after dereddening (\Av=10 mag) and assuming Case B recombination for \Te=\ten{4}K and \Ne=\ten{3} cm$^{-3}$, we find a value of 4\xten{51} 
photons s$^{-1}$.  This is a small fraction of the total expected emission 
from the AGN.  Note that the nucleus has an extinction of up to 70mag along 
the line of sight, estimated from X-ray observations.  However, while being 
within the ionization cone the disk is well outside the obscuring torus, and
subject to a a much lower extinction from the nucleus.  We
thus assume \Av=10 mag as a reasonable upper estimate of the foreground
extinction, following Schreier et al. (1996)
who found a peak extinction of \Av=7 mag in the optical.

The mass of the gas responsible for both extended and unresolved nuclear
emission is estimated from the standard relation,
$L(\PA) = \Np \Ne V 4\pi\cal J(\PA) = \Ne$4$\pi\cal J(\PA) {\rm M}/{\rm m_H}$: 
\begin{equation}
M = 3.9\xten{3}\Mo\,\,F_{-13}^{\rm obs}(\PA)10^{0.059(\Av-10)}
\left(\frac{\Ne}{\ten{3}}\right)^{-1}
\end{equation}
where \Np\ is the proton density, V is the volume of the emitting gas,
${\cal J}(\PA)$ is the line emissivity,
$\rm m_{\rm H}$ is the proton mass,
$F_{-13}^{\rm obs}(\PA)$ is the observed \PA\ flux
in units of \ten{-13} erg cm$^{-2}$ s$^{-1}$ and $10^{0.059(\Av-10)}$
is the reddening correction if \Av\ is different from the assumed 10 mag.
This modest mass estimate of M $\sim$4\xten{3}\Mo\ 
depends mostly on the assumed gas density (\Ne=\ten{3} cm$^{-3}$)
and is extremely uncertain, but we believe 
that a disk of \ten{3}\Mo \ to \ten{5}\Mo \ is quite feasible.

The \PA\ disk is consistent with being perpendicular to the plane of the 
dust lane, and lying along the major axis of the large elliptical galaxy
(see Fig. \ref{fig:pacont}).
It is now widely accepted that the large gas/dust disk of Cen A (i.e. the dust 
lane) was acquired in a recent merger process, and that differential 
precession has led to the observed warped structure (e.g. \cite{tubbs:80}).   
In the central region of Cen A, the precession time is only \ten{7}\ years,
and the orthogonal alignment of our small disk may be consistent with numerical 
studies of the evolution of gas disks in bulge systems (e.g. \cite{quillen:92}).
If we thus interpret the observed emission as being from the warped 
outer portions of an accretion disk around the active nucleus, we conclude 
that even on the relatively small spatial scale of a few parsecs, the disk 
is dominated by the gravitational potential of the galaxy as a whole and not 
by the symmetry of the AGN and its jet.  

For a non-rotating black hole (BH), one would expect the jet to have its 
direction determined by the angular momentum axis of the inner gaseous disk, 
while for a Kerr black hole, it would be expected to be aligned along the 
spin axis of the black hole itself (\cite{bardeen:75}).  Close to a Kerr BH, 
($r\simeq$\,Gm/c$^2$\,$\sim$\ten{14}\,cm $\sim$\ten{-4}\, pc), the 
disk itself will warp to become normal to the spin axis.  In our  
Cen A data, on a few parsec scale, the normal to the \PA\ disk and
the radio jet are misaligned by $\sim$ 70\arcdeg, in projection. 
We conclude that 
if Cen A contains a rotating black hole, then either the gas disk must be 
outside the sphere of influence of the black hole, or it was formed recently 
enough that it has not yet become aligned with the spin axis; for a 
non-rotating black hole, we find that the disk must become significantly 
warped away from being normal to the jet inside a radius of $\sim$2pc.

\cite{pringle:97}
has recently modeled self-induced warping of accretion disks in AGNs.  He 
finds that disks are likely to be warped at $R > 0.02$pc around 
a $\sim$\ten{8}\Mo\ BH.
Our results show a constant position angle for the disk in to $\sim$ 2pc,
suggesting the somewhat weak upper limit of $\sim$ \ten{10}\Mo\
for the mass of the black hole. 

\section{\PA\ Features related to the X-ray/Radio Jet}

\begin{figure}
\centering
\epsfig{file=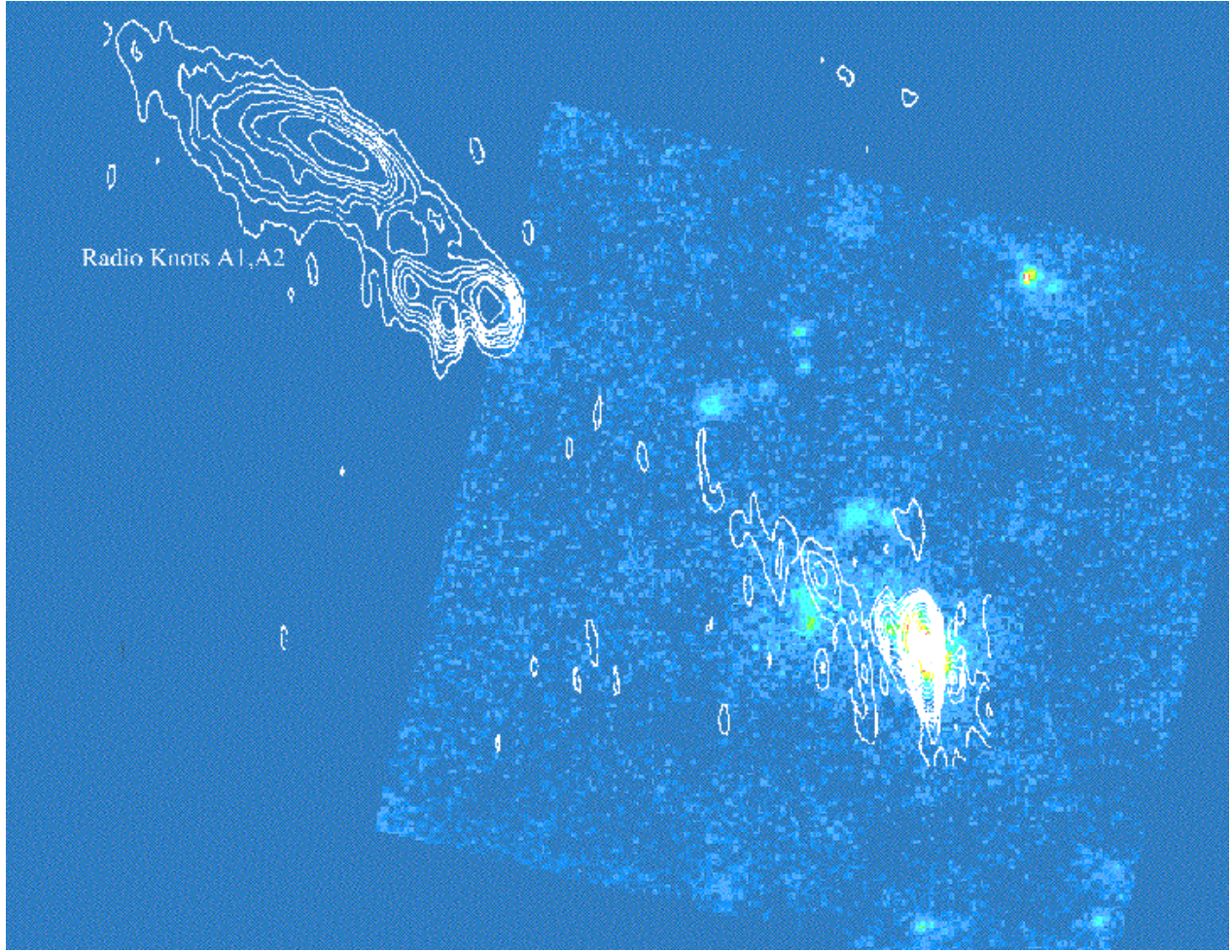,width=0.7\linewidth,angle=-90}
\vskip 0.5cm
\caption{\label{fig:radio} 6cm isophotes of the nuclear region of NGC 5128  
(Feigelson, private communication; see also \cite{clarke:86}) overlayed on 
our \PA\ image.}
\end{figure}

Three of the relatively strong \PA\ features detected appear spatially 
associated with the X-ray/radio jet of Centaurus A.
In Figure \ref{fig:radio} 
we overlay the \PA\ image with a 6 cm radio map 
(Feigelson, private communication; see also Figure 1 of \cite{clarke:86}).  
We align the two images assuming that the radio and NIR nuclei are coincident 
and note that the extended N-S extent of the radio contours 
is the result of the $0\farcs3\times1\farcs1$ radio beam.

The kiloparsec scale X-ray/radio jet has an average
PA = 55\arcdeg$\pm$7\arcdeg,
the innermost knots have a slightly larger PA$\simeq$60\arcdeg\ and
the VLBI milliarcsecond jet and counterjet (cf. \cite{jones:96}) are at 
PA$\simeq$51\arcdeg$\pm$3\arcdeg.    
We note that the multiple components of the radio jet within a few 
arcseconds of the nucleus were not resolved by 
\cite{burns:83} and were reported as N1.
They are clearly visible in the contour map (Fig. \ref{fig:radio})
and Figure 1 of
Clarke et al. (1986), as part of their 6\arcsec\ nuclear jet,
although the individual 
components were not discussed.  We label the compact, innermost knot N1a and 
see that it corresponds positionally with the \PA\ ``finger''$\simeq$0\farcs9 
from the nucleus at PA $\simeq64\arcdeg\pm11\arcdeg$
(see also the small inset in Fig. \ref{fig:pacont}).  The two \PA\ arcs A 
and B, at distances of $\simeq$2\farcs9 and $\simeq$3\farcs5 from the nucleus, 
each extending over $\simeq20$ pc, are seen to lie $\simeq$ 15 and
30 pc, respectively, on either side of another knot of the nuclear jet.
These may result from an outgoing shock created by 
interaction between the radio emitting nuclear ejecta and a gas cloud.  We 
note that the radio jet has a steep spectrum at this distance from the nucleus 
(\cite{clarke:86}),
suggesting that the gas is still too hot along the jet 
axis to show line emission.  We can expect, with adequate resolution, 
to see X-ray emission from this region of the jet.

\section{Summary}

We summarize the key results of our 2.2 $\mu$m continuum and \PA\ 
observations of the inner region of NGC 5128 as follows:

1) We see an extended continuum source with the rather 
regular and smooth structure expected for an elliptical galaxy.  
This smoothness contrasts strongly with the filamentary structure observed 
in the WF/PC I-band image.    

2) We identify a strong unresolved ($r<$0\farcs1) central source as the 
nucleus of the galaxy.  Its position is consistent with 
ground-based IR and radio data and the peak of reddening and 
polarization found with WF/PC-1.  The nucleus has an intensity of 
(2.3$\pm$0.1)\xten{-15} erg cm$^{-2}$ s$^{-1}$ \AA$^{-1}$ based on our 
continuum observations, and a flux of 
(5.7$\pm$0.6)\xten{-14} erg cm$^{-2}$ s$^{-1}$ in \PA.  

3) We see a prominent elongated structure in \PA\, centered on the nucleus, 
with a major to minor axis ratio of $\sim2$, extending $\simeq2$\arcsec along position angle $\simeq$ 33\arcdeg, 
perpendicular to the dust lane.  We interpret this as an inclined $\sim 40$ parsec diameter thin 
nuclear disk of ionized gas.

4) We see several smaller \PA\ features, including three relatively 
strong ones, which appear spatially related to the X-ray/radio 
jet.  We interpret them as circumnuclear gas clouds shocked by the jet.

The disk, with a radius of $\sim 1\farcs2$, corresponding to $\sim20$pc 
at 3.5 Mpc, is among the smallest ever observed at the 
nucleus of an AGN in the optical/near-IR.  It is significantly smaller than 
the hundred or more parsec stellar disks presumed to exist at the centers of 
many galaxies.  Even 
at this small scale, the disk is not perpendicular to the jet.
It is, however, consistent with being perpendicular to the dust lane 
and oriented along the major axis of the bulge.
If it represents the warped outer portion of an accretion disk around the 
active nucleus, we conclude that even a few parsecs from the nucleus,  
the disk is dominated by the galaxy gravitational potential 
and not by the symmetry of the AGN and its jet. 

Further NICMOS observations are planned in 2$\mu$m polarized light   
and [Fe\,II] to provide emission mechanism diagnostics and determine 
the geometry of the nuclear radiation field.
If a warped disk illuminated by the nucleus is the correct model, systematic 
changes in the polarization angle as a function of disk position should be 
seen.  High spatial resolution infrared spectroscopy is planned to measure 
gas kinematics in the disk, and thus determine the mass of the black hole.

\acknowledgements

A.M. and N.C. acknowledge support through GO grants G005.44800,
G005.76700, G005.70200 from
Space Telescope Science Institute, which is operated by the Association
of Universities for Research in Astronomy, Inc., under NASA contract
NAS 5--26555.
We thank Howard Bushouse, Alex Storrs, John Mackenty, Chris Skinner,
Eddy Bergeron and other STScI NICMOS staff for 
invaluable assistance in understanding the NICMOS instrument and its 
calibration.  We thank Nino Panagia, Martino Romaniello, Ernesto Oliva, 
George Miley, and Gary Bower for useful comments and discussions.  We 
thank Eric Feigelson for providing an unpublished radio map
and Zolt Levay for help in producing plates for this paper.

\begin{thebibliography}{}
%
\bibitem[Antonucci 1993]{antonucci:93}%
Antonucci R.R.J., 1993, \araa, 31, 473
%
\bibitem[Bailey et al. 1986]{bailey:86}%
Bailey J., Sparks W.B., Hough J.H., Axon D.J., 1986, Nature, 322, 150
%
\bibitem[Bardeen \& Petterson 1975]{bardeen:75}%
Bardeen J.M and Petterson J.A., 1975, \apjl, 195, 65
%
\bibitem[Blandford 1991]{blandford:91}%
Blandford R.D., 1991, ``Physics of AGN'', Proceedings of Heidelberg Conference,
Springer-Verlag, eds. W.J. Duschl and S.J. Wagner, p. 3
%
\bibitem[Burns et al. 1983]{burns:83}%
Burns J.O., Feigelson E.D., Schreier E.J., 1983, \apj, 273, 128
%
\bibitem[Bower et al. 1998]{bower:98}%
Bower G.A., Green R.F., the Space Telescope Imaging Spectrograph 
Investigation Definition Team, et al., 1998, \apjl, in press
%
\bibitem[Bushouse et al. 1997]{bushouse:97}%
Bushouse H., Skinner C.J., MacKenty J.W, 1997,
NICMOS Instrument Science Report, 97-28 (Baltimore STScI)
%
\bibitem[Cardelli, Clayton and Mathis (1989)]{reddenlaw}%
Cardelli J.A., Clayton G.C. and Mathis J.S., 1989, \apj, 345, 245
%
\bibitem[Clarke et al. 1986]{clarke:86}%
Clarke, D.A., Burns, J.O., and Feigelson, E.D., 1986,\apj, 300, L41
%
\bibitem[Dobereiner et al. 1996]{dobereiner:96}
Dobereiner S., Junkes N., Wagner S.J., Zinnecker H., Fosbury R., 
Fabbiano G., Schreier E.J., 1996, \apj, 470, L15
%
\bibitem[Dufour et al. 1979]{dufour:79}
Dufour R.J., van den Bergh S., Harvel C.A., Martins D.H., Schiffer F.H., 
Talbot R.J., Talent D.L., Wells D.C., 1979, \aj, 84, 284
%
\bibitem[Feigelson et al. 1981]{feigelson:81}%
Feigelson E.D., Schreier E.J., Delvaille J.P., Giacconi R.,
Grindlay J.E., Lightman A.P., 1981, \apj, 251, 31
%
\bibitem[Ferrarese et al. 1996]{ferrarese:96}
Ferrarese L., Ford H.C., Jaffe W., 1996, AJ, 470, 444
%
\bibitem[Harms et al. 1994]{harms:94}%
Harms R.J., Ford H.C., Tsvetanov Z.I., Hartig G.F., Dressel L.L.,
Kriss G.A., Bohlin R.C., Davidsen A.F., Margon B., Kochhar A.K.,
1994, \apj, 435, L35
%
\bibitem[Hui et al. 1993]{hui:93}
Hui X., Ford H.C., Ciardullo R., Jacoby G.H. 1993, \apj, 414, 463
%
\bibitem[Krist \& Hook 1997]{tinytim}%
Krist J.E. and Hook R., 1997, TinyTim User Guide, version 4.4 (Baltimore STScI)
%
\bibitem[Jones et al. 1996]{jones:96}%
Jones D.L., Tingay S.J., Murphy D.W., et al., 1996, \apj, 466, L63
%
\bibitem[MacKenty et al. 1997]{mackenty:97}%
MacKenty J.W., et al., 1997, NICMOS Instrument Handbook, Version 2.0 
(Baltimore STScI)
%
\bibitem[Macchetto et al. 1997]{macchetto:97}%
Macchetto F.D., Marconi A., Axon D.J.,  Capetti A., Sparks W.B., Crane P., 1997, \apj 489, 579
%
%
\bibitem[Osterbrock (1989)]{osterbrock:89}%
Osterbrock D.E., 1989, Astrophysics of Gaseous Nebulae and Active Galactic 
Nuclei, University Science Books, p. 146
%
\bibitem[Packham et al. 1996]{packham:96}%
Packham C., Hough J.H., Young S., Chrysostomou A., Bailey J.A., 
Axon D.J., Ward M.J., 1996, \mnras, 278,406
%
\bibitem[Pringle (1997)]{pringle:97}%
Pringle J.E., 1997, \mnras, 292, 136
%
\bibitem[Quillen et al. 1992]{quillen:92}%
Quillen A.C., de Zeeuw P.T., Phinney E.S., T.G. Phillips, 1992, \apj, 391, 121
%
\bibitem[Robinson 1997]{robinson:97}%
Robinson A., 1997, ASP Conference Series, Eds. B.M. Peterson,
F.-Z. Cheng and A.S. Wilson, Vol. 113, p. 280
%
\bibitem[Rydbeck et al. 1993]{rydbeck:93}%
Rydbeck G., Wiklind T., Cameron M., Wild W., Eckart A., Genzel R., 
Rothermel H., 1993, \aap, 270, L13
%
\bibitem[Schreier et al. 1979]{schreier:79}%
Schreier E.J., Feigelson E., Delvaille J., Giacconi R., Grindlay J.,
Schwartz D.A., Fabian A.C., 1979, \apj, 234, L39
%
\bibitem[Schreier et al. 1981]{schreier:81}%
Schreier E., Burns J.O., Feigelson E.D., 1981,\apj, 251, 523
%
\bibitem[Schreier et al. 1996]{schreier:96}%
Schreier E.J., Capetti A., Macchetto F., Sparks W.B., Ford H.C., 
1996, \apj, 459, 535
%
\bibitem[Schreier et al. 1998]{schreier:98}%
Schreier E.J., Marconi, A., Capetti A., Caon, N., Axon, D., Macchetto F.,
in preparation.
%
\bibitem[Skinner at al. 1997]{pedestal}%
Skinner C.J., Bergeron L.E., Daou D., 1997, HST Calibration Workshop,
Eds. S. Casertano et al. (Baltimore STScI), in press
%
\bibitem[Tubbs 1980]{tubbs:80}%
Tubbs A.D., 1980, \apj, 241, 969
%
\bibitem[Vigroux 1997]{vigroux:97}%
Vigroux, IAU Symposium 186, 1997
%

\end{thebibliography}
\end{document}